\documentclass[letterpaper,12pt]{article}
\usepackage{tabularx} 
\usepackage{amsmath}
\usepackage{amsmath,bm}

\usepackage{graphicx} 
\usepackage[margin=1in,letterpaper]{geometry} 
\usepackage{cite} 
\usepackage[final]{hyperref} 
\hypersetup{
	colorlinks=true,       
	linkcolor=blue,        
	citecolor=blue,        
	filecolor=magenta,     
	urlcolor=blue         
}
\usepackage{amsfonts}
\usepackage{bbold}

\begin{document}
\title{Unitary and non-unitary operations on the Poincaré sphere and Pancharatnam-Berry phase with $\mathbf{Z}$ matrices}
\author{M. A. Kuntman,\; A. Kuntman and E. Kuntman}

\maketitle

\begin{abstract}

In polarization optics unitary and non-unitary operations can be carried out by the Jones matrix. $\mathbf{Z}$ matrix is the $4\times 4$ analogue of the Jones matrix and the Mueller matrix of a nondepolarizing optical medium can be written as $\mathbf{M}=\mathbf{Z}\mathbf{Z}^*$. Jones matrix acts on the two component complex Jones vector, while the 
$\mathbf{Z}$ matrix acts on the four component real Stokes vector. Polarizer and retarder $\mathbf{Z}$ matrices can be written in compact forms in terms of the components of the position vector on the Poincaré sphere. In this note it is shown that the Pancharatnam-Berry geometric phase can be demonstrated by unitary and non-unitary $\mathbf{Z}$ matrix operations.   
\end{abstract} 

\section{Introduction}

If the optical medium is deterministic its optical properties can be represented by a Jones matrix, and alternatively, by a vector state $|h\rangle$ or by a matrix state $\mathbf{Z}$ \cite{KKA, KKPA}. Components of the vector state can be parametrized as:




\begin{equation}
|h\rangle=\begin{pmatrix}
 \tau\\\alpha\\\beta\\\gamma
\end{pmatrix},
\end{equation}
where $\tau$, $\alpha$, $\beta$ and $\gamma$ are generally complex numbers. If the overall phase is not taken into account $\tau$ can be chosen to be real and positive. In that case, the number of independent real parameters will be seven. 
It was shown that $\alpha$, $\beta$ and $\gamma$ can be associated with complex anisotropy parameters of the optical system\cite{KKA}. In terms of spectroscopic parameters, isotropic phase retardation ($\eta$), isotropic amplitude absorption ($\kappa$),  circular dichroism (CD), circular birefringence (CB), horizontal and 45$^{\circ}$ linear dichorism (LD and LD'), horizontal and 45$^{\circ}$ linear birefringence (LB and LB'),
$\tau$, $\alpha$, $\beta$ and $\gamma$ can be written as:
\begin{align}
\tau&=e^{-\frac{i\chi}{2}}\cos\left(\frac{T}{2}\right)\,
&\alpha&=-e^{-\frac{i\chi}{2}}\frac{iL}{T}\sin\left(\frac{T}{2}\right)\\
\beta&=-e^{-\frac{i\chi}{2}}\frac{iL'}{T}\sin\left(\frac{T}{2}\right)\, &\gamma&=e^{-\frac{i\chi}{2}}\frac{iC}{T}\sin\left(\frac{T}{2}\right)
\end{align}
where $\chi=\eta-i\kappa$, $L=LB-iLD$, $L'=LB'-iLD'$,  $C=CB-iCD$,  $T=\sqrt[]{L^2 + L'^2 + C^2}$.

$|h\rangle$ vector is the generator of the nondepolarizing Mueller matrix, and
$\mathbf{Z}$ is the matrix form of $|h\rangle$:

\begin{equation}
\mathbf{Z}=\begin{pmatrix}
\tau&\alpha&\beta&\gamma\\
\alpha&\tau&-i\gamma&i\beta\\
\beta&i\gamma&\tau&-i\alpha\\
\gamma&-i\beta&i\alpha&\tau
\end{pmatrix}.
\end{equation}
In terms of $\mathbf{Z}$ matrices Mueller matrix of a deterministic optical system can be written as
\begin{equation}\label{ZZ=M}
\mathbf{M}= \mathbf{Z}\mathbf{Z^*}=\mathbf{Z^*}\mathbf{Z}.
\end{equation}

$\mathbf{Z}$ matrix is a $4\times 4$ analogue of the Jones matrix, $\mathbf{J}$ \cite{KKCA}:
\begin{equation}
\mathbf{J}=\begin{pmatrix}
\tau+\alpha&\beta-i\gamma\\\beta+i\gamma&\tau-\alpha
\end{pmatrix}.
\end{equation}

Jones matrix transforms two component Jones vector, $|E\rangle$, into Jones vector, $|E'\rangle$, and Mueller matrix transforms  four component Stokes vector, $|s\rangle$, into Stokes vector, $|s'\rangle$:
\begin{equation}
|E'\rangle=\mathbf{J}|E\rangle;\quad |s'\rangle=\mathbf{M}|s\rangle,
\end{equation}
where $|s\rangle= (s_0,s_1,s_2,s_3)^T$, \: $|s'\rangle= (s'_0,s'_1,s'_2,s'_3)^T$ ($s_i$ and $s'_i$ are real numbers). 

$\mathbf{Z}$ matrix transforms the Stokes vector of completely polarized light, $|s\rangle$, into $|\Tilde{s}\rangle$ vector:
\begin{equation}
|\Tilde{s}\rangle=\mathbf{Z}|s\rangle,
\end{equation}
where $|\Tilde{s}\rangle=(\Tilde{s}_0,\Tilde{s}_1,\Tilde{s}_2,\Tilde{s}_3)^T$ and $\Tilde{s}_0, \Tilde{s}_1, \Tilde{s}_2$ and $\Tilde{s}_3$ are, in general, complex numbers. It can be shown that, just like the Jones matrix, the $\mathbf{Z}$ matrix and hence the $|\Tilde{s}\rangle$ vector bears the total phase introduced by the optical system \cite{KKCA}:
\begin{equation}
\langle E|E'\rangle=\langle E|\mathbf{J}|E\rangle=\frac{1}{2}\langle s|\mathbf{Z}|s\rangle=\frac{1}{2}\langle s|\Tilde{s}\rangle.
\end{equation}
\section{Polarization in spherical coordinates}
The normalized Jones vector, $|E\rangle$, of elliptically polarized light with azimuth $\theta$ and ellipticity $\phi$ is given by

\begin{equation}\label{jones}
|E\rangle=e^{i\chi}\begin{pmatrix} \cos{\theta}\cos{\phi}-i\sin{\theta}\sin{\phi}\\ \sin{\theta}\cos{\phi}+i\cos{\theta}\sin{\phi}
\end{pmatrix},
\end{equation}
where $e^{\chi}$ is the overall phase. 

Polarization state of completely polarized light can be represented as a point on the Poincaré sphere with longitute $2\theta$ and latitude $2\phi$. Points on the equator ($\phi=0$) represent linear polarization. 
Points on the upper (lower) hemisphere correspond to right (left) handed elliptical polarization. The north and south pole of the
sphere represent the right-handed and left-handed circular polarization. Diametrically opposite
points represent pairs of orthogonal polarization.

The Stokes vector, $|s\rangle$, corresponding to the Jones vector given by Eq.\eqref{jones} can be written as 
\begin{equation}
|s\rangle=\begin{pmatrix}s_0\\s_1\\s_2\\s_3\end{pmatrix}=\begin{pmatrix}1\\ \cos{2\theta}\cos{2\phi}\\ \sin{2\theta}\cos{2\phi}\\ \sin{2\phi}\end{pmatrix},
\end{equation}
where $s_0, s_1, s_2$ and $s_3$ are normalized Stokes parameters of complete polarization. The stokes vector is directly related with the position vector $\overrightarrow{r}$ on the Poincare sphere: 
\begin{equation}\label{r}
\overrightarrow{r}=\begin{pmatrix}s_1\\s_2\\s_3\end{pmatrix}=\begin{pmatrix}\cos{2\theta}\cos{2\phi}\\ \sin{2\theta}\cos{2\phi}\\ \sin{2\phi}\end{pmatrix}.
\end{equation}
\section{Hermitian and unitary operations on the Poincaré sphere by $\mathbf{Z}$ matrices}

If $\tau, \alpha, \beta$ and $\gamma$ are real, vector state $|h\rangle$ corresponds to a general diattenuator. If $\tau=1,\: \alpha=\cos{2\theta}\cos{2\phi},\: \beta=\sin{2\theta}\cos{2\phi}$\: and $\gamma=\sin{2\phi}$,\: $|h\rangle$ generates a pure polarizer Mueller matrix that projects any input polarization state onto a polarization state on the Poincare sphere defined by the angles $2\theta$ and $2\phi$:
\begin{equation}
|h\rangle_P=\begin{pmatrix}\tau\\\alpha\\\beta\\\gamma\end{pmatrix}=\begin{pmatrix}1\\\cos{2\theta}\cos{2\phi}\\\sin{2\theta}\cos{2\phi}\\\sin{2\phi}\end{pmatrix}.
\end{equation}
The Mueller that is generated by $|h\rangle_P$ can be written as an outer product of $|h\rangle_P$ with itself
\begin{equation}\mathbf{M}_P=|h\rangle_p\langle h|_p \end{equation}By using this property it can be easily shown that $\mathbf{M}_P$ is a projection operator:\begin{equation}\mathbf{M}_P|s\rangle=\langle h|_P |s \rangle |h \rangle_P\end{equation}

$|h\rangle_P$ corresponds to the Hermitian $\mathbf{Z}_P$ matrix:
\begin{equation}
\mathbf{Z}_P=\begin{pmatrix}1&\cos{2\theta}\cos{2\phi}&\sin{2\theta}\cos{2\phi}&\sin{2\phi}\\\cos{2\theta}\cos{2\phi}&1&-i\sin{2\phi}&i\sin{2\theta}\cos{2\phi}\\\sin{2\theta}\cos{2\phi}&i\sin{2\phi}&1&-i\cos{2\theta}\cos{2\phi}\\\sin{2\phi}&-i\sin{2\theta}\cos{2\phi}&i\cos{2\theta}\cos{2\phi}&1\end{pmatrix}.
\end{equation}
The corresponding Mueller matrix can also be written as
\begin{equation}
\mathbf{M}_P=\mathbf{Z}_p\mathbf{Z}^*_p
\end{equation}

If  $\tau$ is real but $\alpha,\: \beta$ and $\gamma$ are pure imaginary numbers then $|h\rangle$ vector represents a general retarder.
In particular, if $\tau=\cos{\delta/2}$, $\alpha=-i n_x\sin{\delta/2}$, $\beta=-in_y\sin{\delta/2}$ and $\gamma=-in_z\sin{\delta/2}$ the associated $\mathbf{Z}$ matrix is a unitary matrix that rotates the input Stokes vector ccw on the Poincaré sphere through an angle $\delta$ about an axis $\hat{n}=(n_x, n_y, n_z)^T$:
\begin{equation}\label{unitary}
|h\rangle_R=\begin{pmatrix}C\\-in_xS\\-in_yS\\-in_zS\end{pmatrix}\longrightarrow
\mathbf{Z}_R=\begin{pmatrix}C&-in_xS&-in_yS&-in_zS\\-in_xS&C&-n_zS&n_yS\\-in_yS&n_zS&C&-n_xS\\-in_zS&-n_yS&n_xS&C\end{pmatrix},
\end{equation}
where $C=\cos{\delta/2}$, $S=\sin{\delta/2}$ ($-\pi/2\leq\delta/2\leq \pi/2$). 

$\mathbf{Z}_P$ and $\mathbf{Z}_R$ matrices transform the Stokes vector of complete polarization, $|s\rangle$, into a complex vector $|\Tilde{s}\rangle$ which bears the phase introduced by the optical element.
\section{Pancharatnam-Berry geometric phase with $\mathbf{Z}$ matrices}

When the polarization state of light undergoes a series of cyclic operations
and returns to its original state, the final state differs in an additional overall phase factor from the original one. It was pointed out by Pancharatnam that the phase difference is not only due to the dynamical phase from the accumulated path lengths but also a geometric phase \cite{Panch}. Berry \cite{Berry1} introduced the corresponding theory for quantum mechanical state vector and re-derived the Pancharatnam’s geometric phase \cite{Berry2}. If the cyclic path consists of only great circles, additional dynamical phase will not develop, and the geometric part increases by $\pm\Omega/2$, where $\Omega$ is the solid angle that the geodesic path of cyclic polarization operations subtends
on the Poincaré sphere.

\begin{figure}\begin{center}
  \includegraphics[width=200pt]{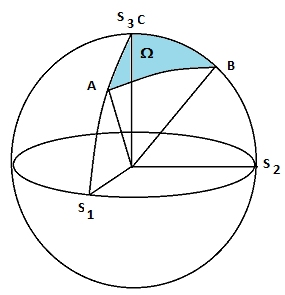}
  \caption{Points on the Poincaré sphere}
  \label{phase}
  \end{center}
\end{figure}

$\mathbf{Z}$ matrices act on the Stokes vectors, and it can be shown that the geometric phase can be calculated by cyclic operations with $\mathbf{Z}$ matrices on the Stokes vectors. For example, let $A$, $B$ and $C$ be three points on the Poincaré sphere represented by the vectors depicted in Fig.\eqref{phase}:
\begin{equation}\overrightarrow{r}_A=\begin{pmatrix}1/2\\0\\ \sqrt{3}/2\end{pmatrix},\quad \overrightarrow{r}_B=\begin{pmatrix}0\\1/2\\ \sqrt{3}/2\end{pmatrix},\quad \overrightarrow{r}_C=\begin{pmatrix}0\\0\\1 \end{pmatrix}.\end{equation} These points correspond to the following Stokes vectors:\begin{equation}|s_A\rangle=\begin{pmatrix}1\\1/2\\0\\ \sqrt{3}/2\end{pmatrix},\quad |s_B\rangle=\begin{pmatrix}1\\0\\1/2\\ \sqrt{3}/2\end{pmatrix},\quad |s_C\rangle=\begin{pmatrix}1\\0\\0\\1 \end{pmatrix}.\end{equation} 

Hermitian cyclic operations can be performed by polarizer $\mathbf{Z}$ matrices that correspond to the projections onto the Stokes vectors $|s_B\rangle$, $|s_C\rangle$ and $|s_A\rangle$ successively:\begin{equation}\mathbf{Z}_P=\mathbf{Z}_A\mathbf{Z}_C\mathbf{Z}_B,\end{equation}where $\mathbf{Z}_P$ is the overall matrix operator that performs the cyclic operation $A\rightarrow B\rightarrow C\rightarrow A$,\begin{equation}\mathbf{Z}_P=\begin{pmatrix}1&1/2&0&\sqrt{3}/2\\1/2&1&-\sqrt{3}i/2&0\\0&\sqrt{3}i/2&1&-i/2\\\sqrt{3}/2&0&i/2&1\end{pmatrix}\begin{pmatrix}1&0&0&1\\0&1&-i&0\\0&i&1&0\\1&0&0&1\end{pmatrix}\begin{pmatrix}1&0&1/2&\sqrt{3}/2\\0&1&-\sqrt{3}i/2&i/2\\1/2&\sqrt{3}i/2&1&0\\\sqrt{3}/2&-i/2&0&1\end{pmatrix}.\end{equation}
The associated geometric phase $\Psi_P$ is
\begin{equation}\Psi_P=\arg(\langle s_A|\mathbf{Z}_P|s_A\rangle)=-4.1066^{\circ}.\end{equation}
Or, equivalently, $\Psi_P$ is the phase of the $Z_{00}$ element of the $\mathbf{Z}_P$ matrix.

Similarly, unitary cyclic operations (rotations) can be performed by retarder $\mathbf{Z}$ matrices. The geodesic path from $A$ to be $B$ is a rotation about the axis $\hat{n}$ defined by the following vector product:
\begin{equation}
\hat{n}=\frac{\overrightarrow{r}_B\times\overrightarrow{r}_A}{||\overrightarrow{r}_B\times\overrightarrow{r}_A||}=\begin{pmatrix}\sqrt{3/7}\\\sqrt{3/7}\\-\sqrt{1/7}\end{pmatrix}.  
\end{equation}
Angle of rotation is the half of the angle $\delta_{AB}$ defined by the scalar product of vectors $\overrightarrow{r}_A$ and $\overrightarrow{r}_B$
\begin{equation}
\delta_{AB}=\arccos{(\overrightarrow{r}_A\cdot\overrightarrow{r}_B)}=\arccos{(3/4)}.
\end{equation}
According to Eq. \ref{unitary}, the unitary operation (cw rotation) on the geodesic from $A$ to $B$ can be performed by the following $\mathbf{Z}_{AB}$ matrix:
\begin{equation}\mathbf{Z}_{AB}=
\begin{pmatrix}C_{AB}&i\sqrt{3/7}S_{AB}&i\sqrt{3/7}S_{AB}&-i\sqrt{1/7}S_{AB}
\\i\sqrt{3/7}S_{AB}&C_{AB}&-\sqrt{1/7}S_{AB}&-\sqrt{3/7}S_{AB}\\i\sqrt{3/7}S_{AB}&\sqrt{1/7}S_{AB}&C_{AB}&\sqrt{3/7}S_{AB}
\\-i\sqrt{1/7}S_{AB}&\sqrt{3/7}S_{AB}&-\sqrt{3/7}S_{AB}&C_{AB}\end{pmatrix},
\end{equation}
where $C_{AB}=\cos{(\delta_{AB}/2)}$, $S_{AB}=\sin{(\delta_{AB}/2)}$. Unitary operation on the geodesic from $B$ to $C$ is a ccw rotation about $\hat{n}=(1,0,0)^T$ through an angle $\delta_{BC}=30^{\circ}$. This operation can be performed by the following $\mathbf{Z}_{BC}$ matrix:
\begin{equation}
\mathbf{Z}_{BC}=\begin{pmatrix}C_{BC}&-iS_{BC}&0&0\\-iS_{BC}&C_{BC}&0&0\\0&0&C_{BC}&-S_{BC}\\0&0&S_{BC}&C_{BC}\end{pmatrix}.
\end{equation}
where $C_{BC}=\cos{(\delta_{BC}/2)}$, $S_{BC}=\sin{(\delta_{BC}/2)}$. The unitary evolution on the geodesic from $C$ to $A$ is a ccw rotation about $\hat{n}=(0,1,0)^T$ through an angle $\delta_{CA}=30^{\circ}$ with the $\mathbf{Z}_{CA}$ matrix, 

\begin{equation}
\mathbf{Z}_{CA}=\frac{1}{2}\begin{pmatrix}C_{CA}&0&-iS_{CA}&0\\0&C_{CA}&0&S_{CA}\\-iS_{CA}&0&C_{CA}&0\\0&-S_{CA}&0&C_{CA}\end{pmatrix}.
\end{equation}
where $C_{CA}=\cos{(\delta_{CA}/2)}$, $S_{CA}=\sin{(\delta_{CA}/2)}$.
As a result, the cyclic unitary operation along the path $A\rightarrow B\rightarrow C\rightarrow A$ can be accomplished by the following $\mathbf{Z}_R$ matrix: 

\begin{equation}\mathbf{Z}_R=\mathbf{Z}_{CA}\mathbf{Z}_{BC}\mathbf{Z}_{AB},\end{equation}

The associated geometric phase $\Psi_R$ is
\begin{equation}\Psi_R=\arg(\langle s_A|\mathbf{Z}_R|s_A\rangle)=-4.1066^{\circ}.\end{equation}
$\Psi_R$ is also the phase accumulated in the $Z_{00}$ element of the $\mathbf{Z}_R$ matrix. The results can be checked by calculating the area subtended by the cyclic path. In this example, the corners of the spherical triangle is defined by the position vectors, hence, it is convenient to use the following formula \cite{Vega}:
\begin{equation}
\frac{1}{2}\Omega=\arctan\left[\frac{\overrightarrow{r}_A\cdot(\overrightarrow{r}_B\times\overrightarrow{r}_C)}{1+\overrightarrow{r}_A\cdot\overrightarrow{r}_B+\overrightarrow{r}_B\cdot\overrightarrow{r}_C+\overrightarrow{r}_C\cdot\overrightarrow{r}_A}\right]=4.1066^{\circ}.
\end{equation}
As a result,
\begin{equation}
\Psi_P=\Psi_R=-\frac{1}{2}\Omega=-4.1066^{\circ}.
\end{equation}


\section{Appendix}
Polarizer state $|h\rangle$ can be written as   
\begin{equation}
|h\rangle_P=\begin{pmatrix}\tau\\\alpha\\\beta\\\gamma\end{pmatrix}=\begin{pmatrix}1\\\cos{2\theta}\cos{2\phi}\\\sin{2\theta}\cos{2\phi}\\\sin{2\phi}\end{pmatrix}.
\end{equation}
Corresponding Jones matrix reads,

\begin{equation}\label{J}
\mathbf{J}=\begin{pmatrix}1+\cos{2\theta}\cos{2\phi}&\sin{2\theta}\cos{2\phi}-i\sin{2\phi}\\\sin{2\theta}\cos{2\phi}+i\sin{2\phi}&1-\cos{2\theta}\cos{2\phi}\end{pmatrix},    
\end{equation}
or, in terms of Pauli vector $\overrightarrow{\sigma}$,
\begin{equation}
\mathbf{J}=I+\overrightarrow{r}\cdot\overrightarrow{\sigma},
\end{equation}
where $I$ is the $2\times 2$ identity, $\overrightarrow{r}$ is the position vector on the Poincare sphere given by Eq.\eqref{r} that corresponds to the transmission axis of the polarizer, and
\begin{equation}
\overrightarrow{\sigma}=\left[\begin{pmatrix}1&0\\0&-1\end{pmatrix},\begin{pmatrix}0&1\\1&0\end{pmatrix},\begin{pmatrix}0&-i\\i&0\end{pmatrix}\right].
\end{equation}

Jones matrix given by Eq.\eqref{J} can also be written as a projection operator in terms of the Jones vector $|E\rangle$ and its Hermitian conjugate:
\begin{equation}
\mathbf{J}=|E\rangle\langle E|,
\end{equation}
where
\begin{equation}
|E\rangle=e^{i\chi}\begin{pmatrix} \cos{\theta}\cos{\phi}-i\sin{\theta}\sin{\phi}\\ \sin{\theta}\cos{\phi}+i\cos{\theta}\sin{\phi}
\end{pmatrix},
\end{equation}
The overall operator for cyclic operations on the Poincaré sphere with polarizer Jones matrices along the path $A\rightarrow B\rightarrow C\rightarrow A$ can be written as
\begin{equation}
\mathbf{J}_P=|E_A\rangle\langle E_A|E_C\rangle\langle E_C|E_B\rangle\langle E_B|.
\end{equation}
Therefore, the geometric phase in terms of projection operators 
\begin{equation}
\Psi_P=\arg(\langle E_A|E_A\rangle\langle E_A|E_C\rangle\langle E_C|E_B\rangle\langle E_B|E_A\rangle)=-4.1066^{\circ}.
\end{equation}


\end{document}